# An evolutionary approach to predict slope displacement of earth embankments under earthquake ground motions

Zhenyang Jin [a],[*], Sanglin Zhao [b], Siyu Fan [c], Hamed Javdanian [d],[*]

[a] *College of Public Security Management, People's Public Security University of China, Beijing 100000, China*
[b] *School of Engineering Management, Hunan University of Finance and Economics, Changsha, Hunan 410000, China*
[c] *School of Cyberspace Security, Beijing University of Posts and Telecommunications, Beijing 100000, China*
[d] *Department of Civil Engineering, Shahrekord University, Shahrekord, Iran*

ABSTRACT

Accurate slope stability analysis of earth embankments under ground shaking is of great importance for practical use in earthquake geotechnics. This study aims to predict soil slope displacements of earth embankments subjected to earthquake loading using evolutionary algorithms. Comprehensive real case histories of slope displacement of earth embankments under past earthquakes in different areas of the world were gathered and analyzed. A robust model was then developed to predict earthquake induced soil slope displacements using gene expression programming (GEP). Characteristics of earthquake ground motion including earthquake magnitude, earthquake predominant period, maximum earthquake acceleration and also geotechnical specifications of earth embankment including yield acceleration and fundamental period of earth embankment were taken as most influential factors on the slope displacements of earth embankments under earthquakes. Subsequently, performance of developed GEP-based predictive model was assessed using a sensitivity analysis under various effective factors. Finally, the accuracy of the predictive model was evaluated through comparison with the available relationships for estimation of seismic soil slope displacements. The results clearly indicate favorable accuracy of developed GEP-based model to predict slope displacements of earth embankments subjected to earthquake ground motions.

## Introduction

Investigating dynamic behavior of earth embankment slopes under earthquake shaking is crucial for applying techniques of landslide disaster prediction and prevention. The seismically induced permanent slope displacement is usually assessed by Newmark [47]'s sliding rigid block (e.g., [5,11,36,48,67]). The sliding mass in the Newmark approach was considered as a rigid block. Several researchers (e.g., [39,56]) studied and modified the sliding block method. Rathje and Antonakos [54] developed an empirical model to predict the seismic displacements of flexible sliding masses. uncertainty of rigid block-based slope displacement results compared to the real cases was examined by Strenk and Wartman [68]. Performance of Newmark-based models were evaluated by Meehan and Vahedifard [44] through case histories describing the slope displacements of earth dams and embankments during past earthquakes [63]. Song et al. [65,66], on the basis of Newmark method, developed multi-block sliding approaches to evaluate the seismic displacement of slopes with multiple potential failure surfaces. Using parametric study, Roy et al. [57] studied the influence of ground motion specifications on the seismic displacement of slopes.

Cho and Rathje [8], using finite element analysis, calculated the slope displacement under shallow crustal earthquakes. Javdanian and Co-workers [27,30,31,45,46,61] studied the slope stability of earth dams using numerical finite element simulation. Seismic stability of layered earth slope is investigated using finite difference numerical modeling [75]. Fotopoulou and Pitilakis [19] compared the calculated earthquake induced slope displacements through numerical analysis with the predicted values by empirical models. Jiao et al. [33] investigated the seismically induced stability of soil slopes using numerical and experimental studies. Some researchers (e.g., [5,73]) studied the seismically behavior of soil slope under earthquake loading using experimental physical modeling. This method can provide real data on dynamic response of soil slope and capture the nonlinearity characteristics of slope materials.

Table 1

Statistical characteristics of slope displacement of earth embankments under earthquakes.

| Statistical | Parameters | | | | | | | |
|---|---|---|---|---|---|---|---|---|
| | $M_w$ | $a_{max}$ (g) | $T_p$ (sec) | $T_d$ (sec) | $a_y$ (g) | $a_y/a_{max}$ | $T_d/T_p$ | D (m) |
| **All data** | | | | | | | | |
| Min | 4.9 | 0.06 | 0.25 | 0.05 | 0 | 0 | 0.117 | 0.001 |
| Max | 8.3 | 0.9 | 0.7 | 1.58 | 0.55 | 3.5 | 4 | 7.696 |
| Average | 7.091 | 0.302 | 0.377 | 0.519 | 0.17 | 0.770 | 1.435 | 1.084 |
| SD | 0.670 | 0.177 | 0.121 | 0.398 | 0.115 | 0.704 | 1.032 | 1.811 |
| **Training** | | | | | | | | |
| Min | 4.9 | 0.06 | 0.25 | 0.05 | 0 | 0 | 0.117 | 0.001 |
| Max | 8.2 | 0.9 | 0.7 | 1.58 | 0.55 | 3.5 | 4 | 7.696 |
| Average | 7.086 | 0.301 | 0.382 | 0.522 | 0.171 | 0.797 | 1.443 | 1.037 |
| SD | 0.702 | 0.181 | 0.125 | 0.396 | 0.118 | 0.735 | 1.049 | 1.787 |
| **Testing** | | | | | | | | |
| Min | 5.5 | 0.07 | 0.25 | 0.05 | 0 | 0 | 0.15 | 0.001 |
| Max | 8.3 | 0.7 | 0.65 | 1.58 | 0.37 | 2.857 | 3.857 | 6.061 |
| Average | 7.105 | 0.303 | 0.365 | 0.512 | 0.168 | 0.691 | 1.415 | 1.220 |
| SD | 0.586 | 0.171 | 0.111 | 0.411 | 0.106 | 0.615 | 1.008 | 1.916 |

Table 2

Pearson correlation coefficients between parameters.

| | $M_w$ | $a_{max}$ (g) | $T_p$ (sec) | $T_d$ (sec) | $a_y$ (g) | $a_y/a_{max}$ | $T_d/T_p$ | D (m) |
|---|---|---|---|---|---|---|---|---|
| $M_w$ | 1 | | | | | | | |
| $a_{max}$ (g) | 0.287 | 1 | | | | | | |
| $T_p$ (sec) | 0.641 | 0.009 | 1 | | | | | |
| $T_d$ (sec) | −0.057 | −0.236 | 0.223 | 1 | | | | |
| $a_y$ (g) | 0.044 | 0.063 | 0.061 | 0.323 | 1 | | | |
| $a_y/a_{max}$ | −0.184 | −0.506 | −0.007 | 0.427 | 0.635 | 1 | | |
| $T_d/T_p$ | −0.394 | −0.253 | −0.181 | 0.869 | 0.260 | 0.446 | 1 | |
| D (m) | 0.330 | 0.244 | 0.274 | −0.292 | −0.375 | −0.366 | −0.348 | 1 |

Within the past years, novel manifestation of modeling, optimization, and issue understanding have been developed as respect the unavoidable advance in computational methods. These perspectives are alluded as soft computing strategies which are exceptionally capable approaches for nonlinear and multivariate modeling [17,28,40]. These demonstrate that the advanced computational algorithms ought to be utilized to precisely evaluate the behavior of earth structures as one of the serious and complex problems in geotechnics [43,58,77]. Some scholars used intelligence systems to predict soil slopes safety factor under seismic loading (e.g., [15,16,71]). Javdanian and Pradhan [29] studied the slope deformation of earth dams subjected to earthquakes using two soft computing techniques of radial basis and feed forward back propagation methods. The study of Huang et al. [21] confirms the matureness and sufficiency of computational algorithms to determine the dynamic response of soil and rock slope systems. They employed the results of large scale shake table experiments. On the basis of numerical database, Cho et al. [7] developed artificial intelligence network based models for prediction of seismically slope displacement. They compared the results of neural based models with the classical regression relationships and indicated high capability of computational approaches in assessment soil slope performance subjected to seismic loading. A neural based model was proposed by Wang and Wu [74] to estimate earthquake induced displacements of flexible and rigid slopes. Using slope displacement data calculated by Newmark rigid block method, Cheng et al. [6] developed a neural network model to predict seismic displacements. They demonstrated the effectiveness of developed models by applying them in the probabilistic risk analysis of slope displacement. Lin et al. [38] studied stability of multilayer earth slope using convolutional neural network. Assessment of susceptibility of earthquake induced landslide by using soft computing techniques are also illustrating robustness of the computational algorithms in dynamic analyses of earth structures [37,78].

In this research, using evolutionary algorithms, a predictive model was developed for calculation of slope displacement of earth embankments under earthquake ground motions. Wide-ranging real case histories of earth embankments under past earthquake from different region of the world were compiled. The data was analyzed and influential parameters that affect the seismic behavior of earth embankments were determined. A predictive model was developed using gene expression programming (GEP) to evaluate earthquake excitation induced slope displacements of earth embankments. The precision of the GEP-based model was assessed. Then, a sensitivity analyses was performed to check the performance of developed predictive model under variation of the influential factors. Finally, the accuracy of proposed GEP-based model in estimation of slope displacement of earth embankments subjected to earthquake ground motions were compared with the available relationships.

**Earth embankment database**

A comprehensive actual database of slope displacement of earth embankments under earthquake ground motions in different regions of the world was compiled. The seismic displacement results refer to homogeneous/nonhomogeneous earth and rockfill dams, embankments, and natural earth slopes. The database includes earth embankments whose dynamic behaviors were well-documented after earthquake ground motion ([4,9,10,13,42]; Nicholas, [2,3,12,20,49–52,64]). The gathered data contain 85 real cases. The parameters maximum horizontal earthquake acceleration ($a_{max}$), predominant earthquake period ($T_p$) and earthquake magnitude ($M_w$) as specifications of earthquake ground motions and the parameters fundamental period of earth embankment ($T_d$) and yield acceleration ($a_y$) as geotechnical specifications of earth embankment were chosen as important factors that affect the soil slope displacement under earthquakes (D). The fundamental period ($T_d$) of earth embankment was attained from the report of case history, if available. Otherwise, the $T_d$ is calculated as $4H/V_s$ [55]. Where $V_s$ is the shear wave velocity in the earth embankment and H is the earth embankment height. The yield acceleration ($a_y$) was attained from slope stability analysis using pseudo-static method [35]. The $a_y$ is equal to the inertial acceleration that yields a safety factor of one in a soil slope pseudo-static analysis. The $D$ is the move of the soil mass downward aligns the slide surface inclination subjected earthquake excitation.

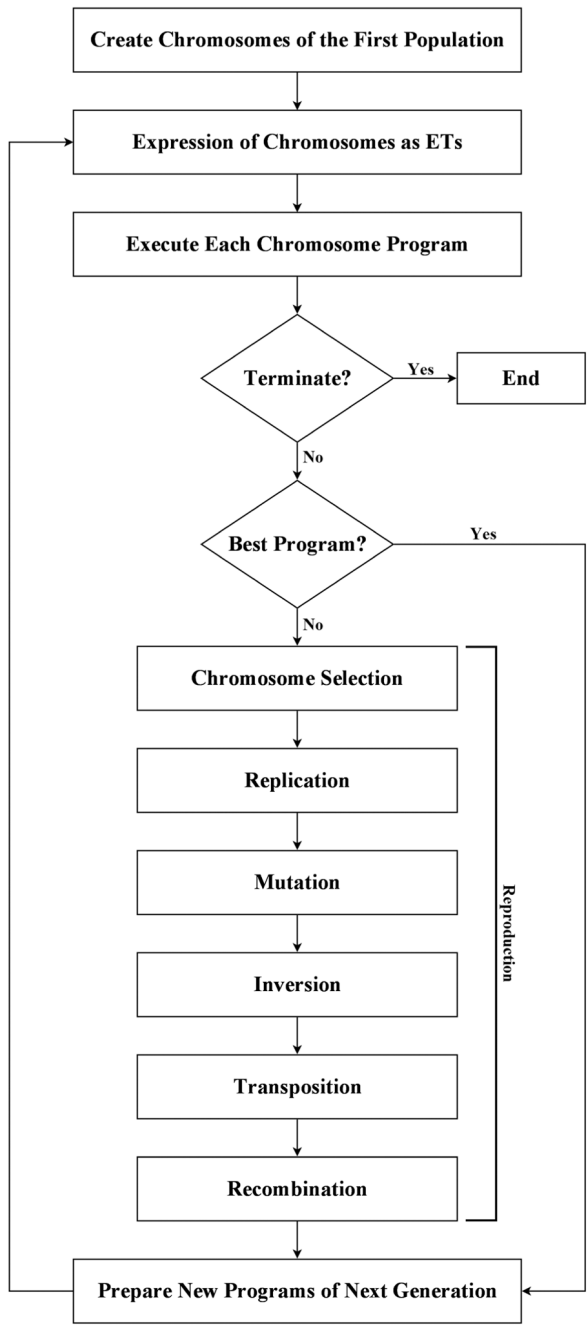

**Fig. 1.** Working procedure of GEP algorithm.

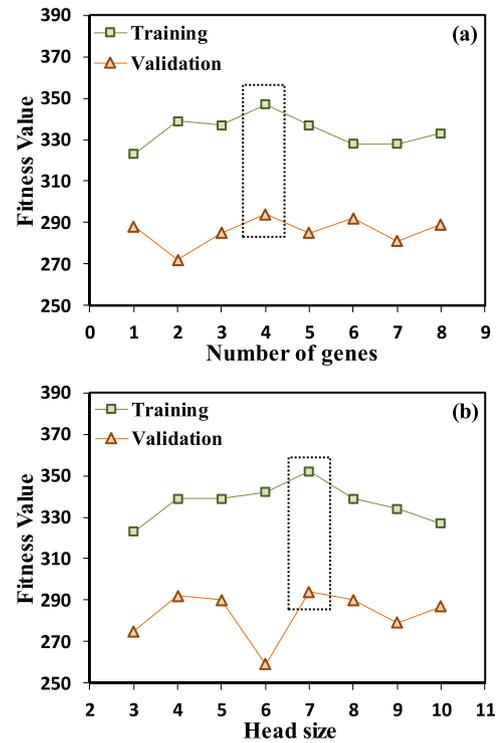

**Fig. 2.** Fitness value variation with, a) number of genes, b) head size.

**Table 3**
Optimal parameters of the GEP-based predictive model.

| Parameter | Value |
| --- | --- |
| Number of chromosomes | 50 |
| Head Size | 7 |
| Number of genes | 4 |
| Linking function | + |
| Function set | +, -, *, / |
| Rate of mutation | 0.0014 |
| Conservative mutation | 0.0037 |
| Permutation | 0.0055 |
| Biased mutation | 0.0055 |
| IS transposition rate | 0.0055 |
| RIS transposition rate | 0.0055 |
| Rate of inversion | 0.0055 |
| Uniform recombination | 0.008 |
| One-point recombination | 0.003 |
| Two-point recombination | 0.003 |
| Rate of gene recombination | 0.003 |
| Rate of gene transposition | 0.003 |

Based on the analyzing collected real cases and reviewing the available studies (e.g., [24,60]) the parameters ratio of fundamental period, $T_d/T_p$, earthquake magnitude, $M_w$, and ratio of yield acceleration, $a_y/a_{max}$, were chosen as input parameters in development of the model. For model development, 75 % of the gathered actual data was applied for training of the model, while 25 % was used for validation stage [25,26]. On the basis of a trial selection method, the data for the stages of training and validation were chosen such that the statistical specifications of both sets as close as feasible [22,62,76]. The statistical specifications of inputs (i.e., $T_d/T_p$, $a_y/a_{max}$, $M_w$) and also output (i.e., $D$) parameters in the training and validation steps and also the all data are introduced in Table 1.

Table 2 presents Pearson correlation coefficient ($r$) between influential parameters and slope displacement of earth embankment under earthquakes, $D$. The correlation is considered statistically significant at the 0.0 level. The results indicated that earthquake induced displacement had a positive relationship with earthquake magnitude, $M_w$, ($r = 0.330$) and negative relationship with parameters of ratio of fundamental period, $T_d/T_p$, ($r = -0.348$) and ratio of yield acceleration, $a_y/a_{max}$, ($r = -0.366$).

**Methodology**

Gene expression programming, GEP, integrates the preponderances of genetic programming, GP, and genetic algorithm, GA, [18]. Set of genes (i.e., chromosomes) are the main elements of GEP and expressed as computational equations that form a chromosome. In the GEP the chromosomes are known as expression trees, ETs, [69]. Each gene has two components including a tail and a head. The tail only includes terminals, while the head includes terminals and operators (e.g., -, +, /, ×) [32]. An appropriate relationship between head length ($L_h$) and tail length ($L_t$) is as following [18]:

$$L_t = L_h(n_{\max} - 1) + 1 \qquad (1)$$

where $L_t$ is the tail length, $L_h$ is the head length, and $n_{max}$ is the maximum operators number.

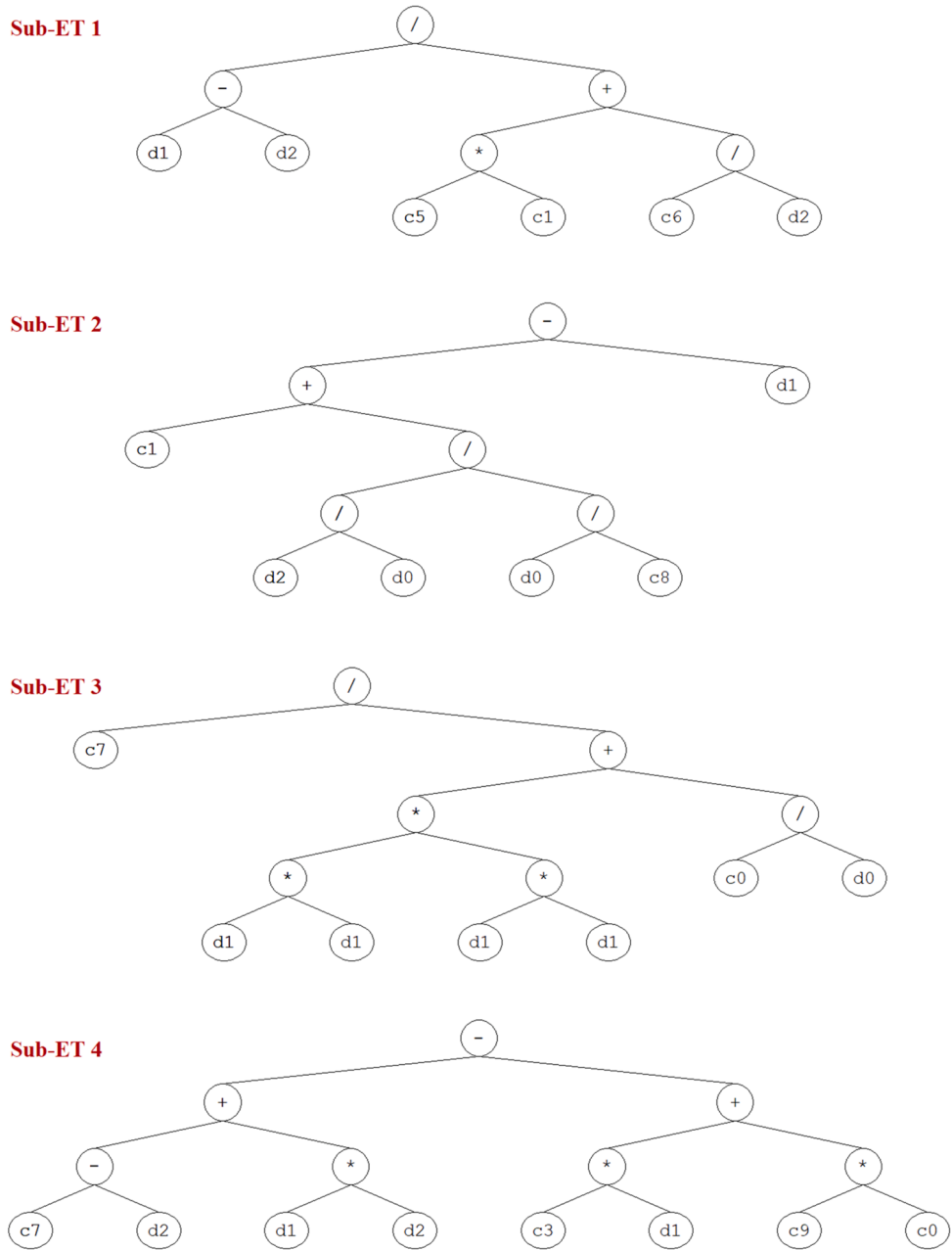

**Fig. 3.** Expression tree of developed GEP-based model.

Genes number and also head length that make up the chromosomes structure should be specified on the basis of complexity of the problem. The number and contribution of operators in the solution process is required to develop a high performance GEP-based model [79]. Generally, number of genes increases with increasing number of parameters of the problem. The functions that connect the genes in the GEP-based models are different for every problem and vary relating to the type of the practical problem. The GEP flowchart is depicted in Fig. 1.

*Model development*

For each problem, it is required to utilize optimal parameters that have an important influence on the performance of developed GEP-based predictive model. In the present research, trial-error technique was utilized to characterize the optimal amounts of GEP parameters, such as head length, chromosome length, gene number, and other setting factors. GEP generates the population by randomly producing individuals from terminals and functions where the structure of chromosome is preparing [72]. The best chromosomes are chosen on the basis of fitness criterion. Selecting an appropriate fitness function is one of the main steps of GEP-based model development and optimization of parameters. One of the most commonly used fitness criterion ($f_c$) to check the robustness of developed GEP-based model is presented as Eq.

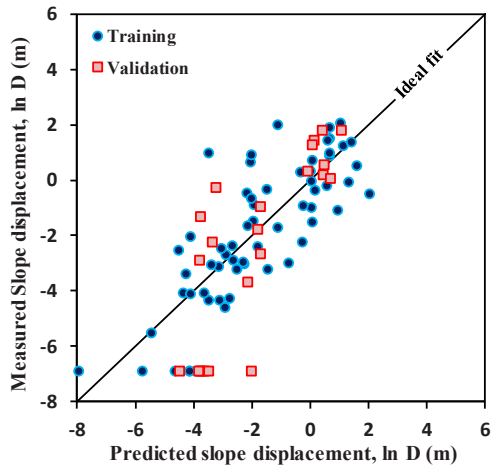

**Fig. 4.** Measured versus GEP-based predicted amounts of slope displacement for training and testing data sets.

**Table 4**
Accuracy of the developed GEP-based model for different stages.

| Stage | Number of data | Error parameters | | | | |
|---|---|---|---|---|---|---|
| | | $R^2$ | MAE | RMSE | SI | Bias |
| Training | 63 | 0.742 | 1.198 | 1.546 | 0.787 | 0.152 |
| Validation | 22 | 0.720 | 1.675 | 2.115 | 0.930 | 0.476 |
| All data | 85 | 0.730 | 1.321 | 1.712 | 0.837 | 0.236 |

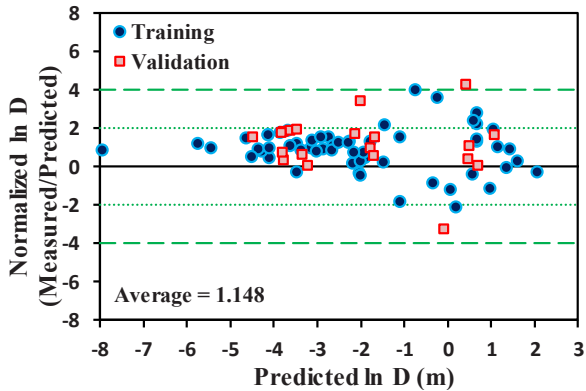

**Fig. 5.** Normalized values of ln D versus GEP-based predicted values of ln D.

(2) [18]:

$$f_c = 1000 \times \left(\frac{1}{1 + RMSE}\right) \quad (2)$$

where the RMSE is root mean square error of individual chromosome which is considered in attaining the regression function. The range of fitness criterion is from 0 to 1000 and 1000 is the ideal fitness. The perfect fit is obtained when RMSE=0 and then $f_c$=1000.

The process of evolutionary is finished on the basis of some convergence criterion. To this end, the number of generations is defined or the process can be finished when multiple generations does not lead to change the best value of fitness. In the present research, a simple mathematical functions set {-, +, /, *} was chosen to demonstrate relationship between influential parameters. Two optimal amounts of gene number and head size were selected based on the variation of fitness function. Fig. 2 depicts variation of fitness values against gene number and head size. The results indicated that the optimal number if genes is 4 and the head size is 7 (Fig. 2a-b).

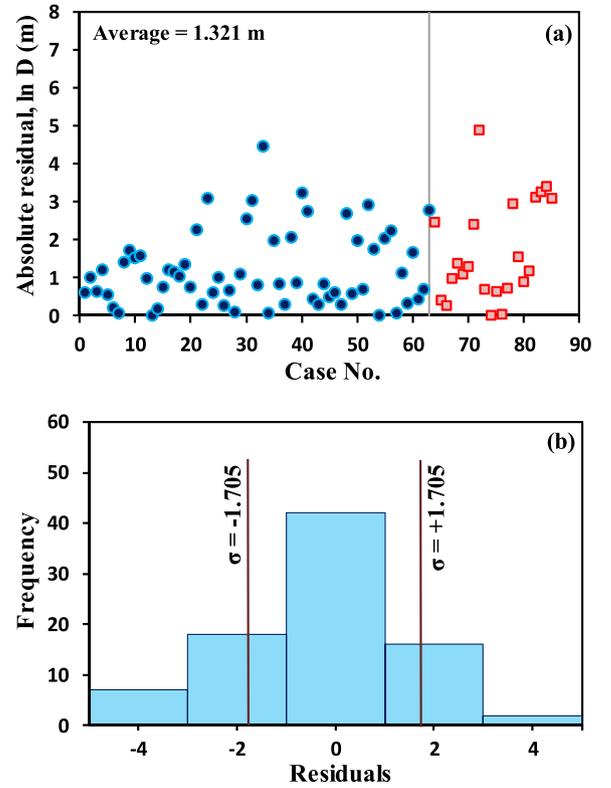

**Fig. 6.** a) residuals values, b) histogram of residuals.

Linking function of addition (+) was applied in many researches (e. g., [14,53,59]), therefore, this function was selected as linking function for developing GEP-based model. The optimal values of other genetic parameters including rate of mutation, conservative mutation, permutation, biased mutation, IS/RIS transposition rate, rate of inversion, uniform recombination, one and two point recombination, rate of gene recombination, and rate of gene transposition are presented in the Table 3. These parameters are utilized to the development of GEP-based predictive model to predict slope displacement of earth embankments subjected to earthquake ground motions.

*Model performance*

Coefficient of determination, $R^2$, root mean square error, RMSE, mean absolute error, MAE, Bias, and scatter index, SI, were utilized to assess the precision of developed GEP-based model for soil slope displacements under earthquake ground motions using Eqs. (3–7):

$$R^2 = 1 - \frac{\sum_{i=1}^{N}(Y_m - Y_p)^2}{\sum_{i=1}^{N}(Y_m - \overline{Y}_m)^2} \quad (3)$$

$$MAE = \frac{1}{N}\left[\frac{\sum_{i=1}^{N}|Y_p - Y_m|}{\sum_{i=1}^{N}Y_m}\right] \quad (4)$$

$$RMSE = \left[\frac{\sum_{i=1}^{N}[Y_p - Y_m]^2}{N}\right]^{0.5} \quad (5)$$

$$SI = \frac{RMSE}{(1/N)\sum_{i=1}^{N}Y_m} \quad (6)$$

$$Bias = \frac{1}{N}\sum_{i=1}^{N}|Y_p - Y_m| \quad (7)$$

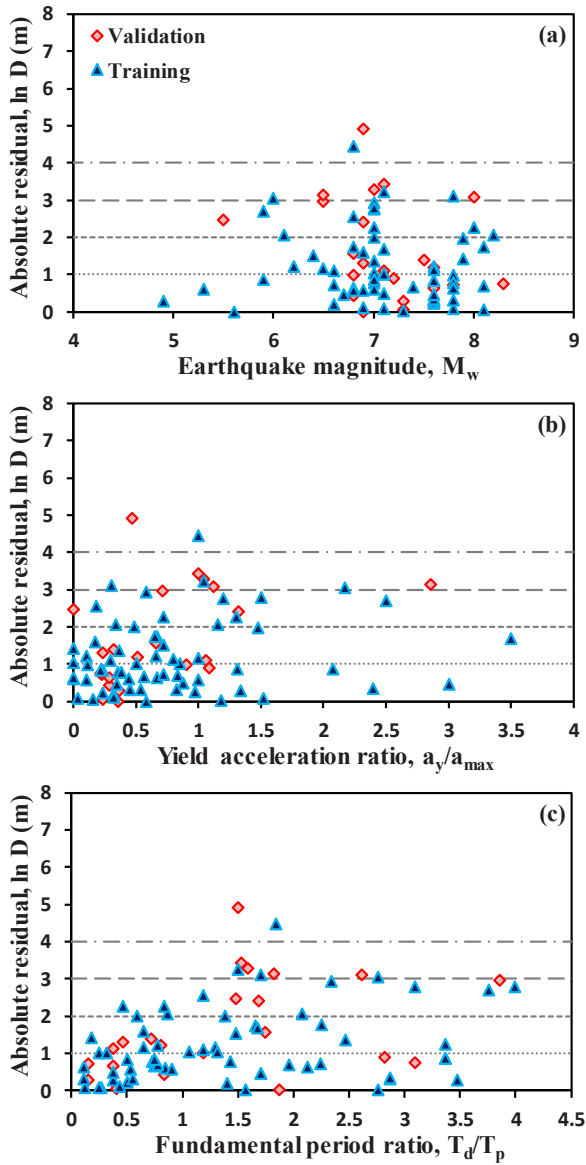

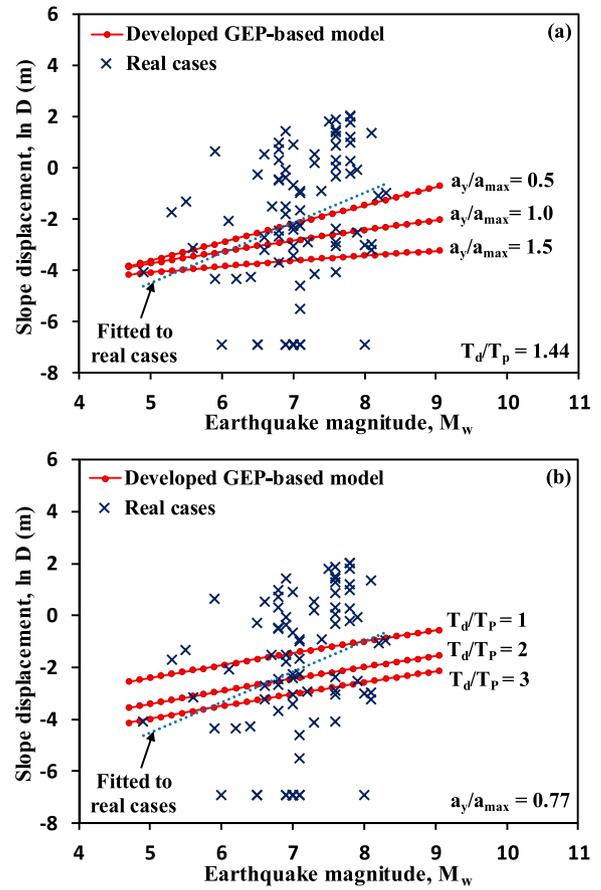

Fig. 8. Variation of GEP-based slope displacement of earth embankments versus earthquake magnitude for different values of, a) ratio of yield acceleration, and b) ratio of fundamental period.

$$\ln D = \frac{6.524\ M_w}{M_w \left(\frac{a_y}{a_{max}}\right)^4 + 7.864} + \frac{\left(\frac{a_y}{a_{max}}\right)\left(\frac{T_d}{T_p}\right) - \left(\frac{T_d}{T_p}\right)^2}{5.55\left(\frac{T_d}{T_p}\right) - 7.052} + \frac{3.647}{M_w^2}$$
$$+ \left(\frac{a_y}{a_{max}}\right)\left(\frac{T_d}{T_p}\right) - \left(\frac{a_y}{a_{max}}\right) - \left(\frac{T_d}{T_p}\right) - 5.098 \qquad (8)$$

Fig. 7. a) The values of residuals versus, a) earthquake magnitude, b) ratio of yield acceleration, and c) ratio of fundamental period.

where, $Y_m$ is the measured earthquake-induced slope displacement, $Y_p$ is the predicted earthquake-induced slope displacement, $\overline{Y}_m$ is the mean of measured earthquake-induced slope displacement, and $N$ is the number of real case histories.

**Results and discussion**

In the present research, several models with various initial parameters were developed. Finally, on the basis of error parameters, the model with the highest precision was selected as the best predictive model. The expression tree of developed GEP-based model including four sub-expression trees (sub-ETs) is depicted in the Fig. 3. The c5, c1, c6, c8, c7, c0, c3 and c9 are constants and d0, d1 and d2 denote $M_w$, $a_y/a_{max}$ and $T_d/T_p$, respectively (Fig. 3). Therefore, the proposed GEP-based model for estimation of soil slope displacement under earthquake ground motion is as Eq. (8):

The accuracy of the developed predictive GEP-based model (Eq. 8) is shown in Fig. 4 by comparing the measured slope displacement under earthquake ground motions versus the amounts predicted by developed model. Results demonstrated that, the amounts of $R^2$, MAE, RMSE, SI and Bias of the GEP-based predictive model for evaluating earthquake-induced slope displacement of earth embankments were respectively 0.742, 1.198, 1.546, 0.787 and 0.152 in the training stage and 0.720, 1.675, 2.115, 0.930 and 0.476 in the validation stage. The amounts of $R^2$, RMSE, MAE, Bias and SI for developed GEP-based model in training and validation stages and also all data sets are presented in Table 4. The results indicate the favorable precision of the proposed model in estimating soil slope displacements of earth embankments under earthquake ground motions.

The Ln D ratio values (i.e., normalized ln D as measured values to the predicted ones) against the predicted values of ln D are illustrated in Fig. 5. This figure illustrates that the average amount of the normalized ln D is 1.148 which confirms that the GEP-based predicted slope displacements were unbiased. For more assessment of the proposed model precision in evaluation of soil slope displacements of earth embankments under earthquakes (D), the residuals (i.e., differences between the predicted and measured amounts) was calculated and shown in Fig. 6a. As demonstrated in this figure, the relative error of the developed predictive GEP-based model is approximately less than ln D = 3 m for both training and validation stages. The analyses results illustrate that the average amount of the absolute residuals is ln D = 1.321 m (Fig. 6a). Fig. 6b also shows the residuals histogram with two lines demonstrating the minus/plus standard deviation of the values of residuals. The residuals values represent the normal distributions with the standard deviations of 1.705 (Fig. 6b).

**Table 5**
Relationships for assessment of earthquake induced displacement of soil slope.

| Relationship | Applied range | Reference |
|---|---|---|
| $\log(D_{(cm)}) = -0.287 - 2.854\left(\frac{a_y}{a_{max}}\right) - 1.733\left(\frac{a_y}{a_{max}}\right)^2 - 0.702\left(\frac{a_y}{a_{max}}\right)^3 - 0.116\left(\frac{a_y}{a_{max}}\right)^4$ | $M_w \leq 8$ $0.01 \leq a_y/a_{max} \leq 0.6$ | Hynes-Griffin and Franklin [23] |
| $\log(D_{(m)}) = 0.9 + \log\left[\left(1 - \frac{a_y}{a_{max}}\right)^{2.53} \times \left(\frac{a_y}{a_{max}}\right)^{-1.09}\right]$ | $0.05 \leq a_y/a_{max} \leq 0.95$ $6.6 \leq M_w \leq 7.2$ | Ambraseys and Menu [1] |
| $\log(D_{(cm)}) = -0.215 + \log\left[\left(1 - \frac{a_y}{a_{max}}\right)^{2.341} \times \left(\frac{a_y}{a_{max}}\right)^{-1.438}\right]$ | $a_y/a_{max} \leq 1$ $0.05 \leq a_y \leq 0.4\,g$ $5.3 \leq M_w \leq 7.6$ | Jibson [34] |
| $\ln(D_{(cm)}) = 5.52 + 0.72\ln(a_{max}) - 4.43\left(\frac{a_y}{a_{max}}\right) - 20.93\left(\frac{a_y}{a_{max}}\right)^2 + 42.61\left(\frac{a_y}{a_{max}}\right)^3 - 28.74\left(\frac{a_y}{a_{max}}\right)^4$ | $0.05 \leq a_y \leq 0.3\,g$ $0.05 \leq a_y/a_{max} \leq 1$ $a_{max} \leq 1\,g$ $4.5 \leq M_w \leq 7.9$ | Saygili and Rathje [60] |
| $\log(D_{(cm)}) = -0.418 - 0.857 \log\left(\frac{a_y}{a_{max}}\right) + 2.26 \log\left(1 - \frac{a_y}{a_{max}}\right)$ | $0.1 \leq a_y/a_{max} \leq 0.9$ | Madiai [41] |
| $\ln(D_{(cm)}) = 6.4 - 8.374\left(\frac{a_y}{a_{max}}\right) - 0.419\left(\frac{a_y}{a_{max}}\right)^2 + 6.366\left(\frac{a_y}{a_{max}}\right)^3 - 7.031\left(\frac{a_y}{a_{max}}\right)^4 + 0.767\ln(a_{max}) + 1.757\ln(T_m)$ | $5.9 \leq M_w \leq 7.6$ $a_{max} \leq 0.3\,g$ | Tsai and Chien [70] |

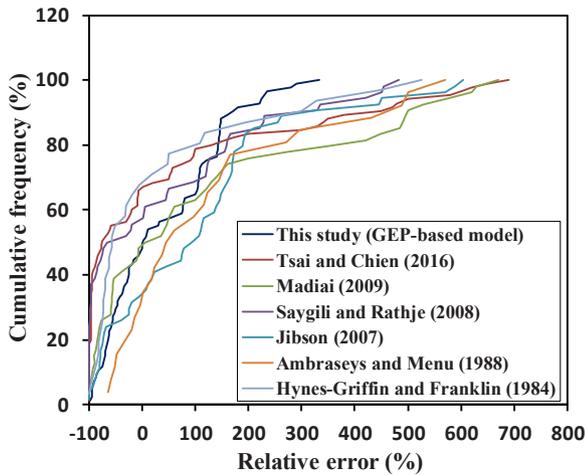

**Fig. 9.** Cumulative frequency distribution versus relative error.

Variation of residuals values versus input parameters (i.e., the ratio of fundamental period, $T_d/T_p$, earthquake magnitude, $M_w$, ratio of yield acceleration, $a_y/a_{max}$) for training and validation data sets are shown in Fig. 7a-c. The results (Figs. 4–7) show the favorable precision of the predictive GEP-based model in assessing soil slope displacement of earth embankments under earthquake ground motions.

*Sensitivity analysis*

In this part, a sensitivity analysis was carried out in order to investigate 1) how each influential parameter affects soil slope displacements under earthquake ground motions and 2) the agreement between the physical behavior of the developed GEP-based model and actual case histories results under different conditions. To this end, the influence of the variation of input parameters on the earthquake induced slope displacements (Ln D) was assessed while the other influential parameters were kept constant at their mean amounts in the case database (Table 1).

The variation of slope displacements of earth embankments predicted by GEP-based predictive model against earthquake magnitude ($M_w$) at different values of $a_y/a_{max}$ is shown in Fig. 8a. Fig. 8b depicts the GEP-based values of Ln D against $M_w$ at different level of $T_d/T_p$. The real values of soil slope displacements of earth embankments under earthquake ground motions and the best fitted curve are also shown in Fig. 8a-b for comparison purpose. As depicted in Fig. 8, the earthquake induced slope displacements increased by increasing earthquake magnitude. Increasing $a_y/a_{max}$ (Fig. 8a) and $T_d/T_p$ (Fig. 8b) led to decrease slope displacements of earth embankments. Generally, comparing variations of ln D against the influential parameters on the earthquake induced soil slope displacements with actual case histories shows the reasonable performance of the proposed GEP-based predictive model for calculation of *D*.

**Comparison with available relationships**

Performance of developed GEP-based predictive model in comparison to the well-known relationships (Table 5) for assessment of soil slope displacement under earthquake ground motion is demonstrated in Fig. 9. These relationships (Table 5) are developed on the basis of Newmark's rigid-block sliding theory. Hynes-Griffin and Franklin [23], Ambraseys and Menu [1], Jibson [34], and Madiai [41] proposed their relationships in terms of yield acceleration to maximum ground acceleration, $a_y/a_{max}$. Saygili and Rathje [60] proposed an equation to estimate earthquake induced soil slope displacements based on the parameters $a_{max}$ and $a_y/a_{max}$. Tsai and Chien [70] considered the effect of mean period of ground motions ($T_m$) in addition to $a_{max}$ and $a_y/a_{max}$. Applied range for each relationship is presented in Table 5. The relative error values of previous relationships were computed for the applied ranges.

Fig. 9 depicts comparison of cumulative frequency of the relative errors for proposed predictive GEP-based model (Eq. 8) and available relationships. The relative error ($E_R$) values was calculated using Eq. (9):

$$E_R = \frac{D_{predicted} - D_{measured}}{D_{measured}} \times 100 \tag{9}$$

where, the $D_{predicted}$ is the earthquake induced slope displacement predicted by the developed GEP-based model and also available recommendations and $D_{measured}$ is the real soil slope displacement under earthquakes.

The relative error values against the measured values of slope displacements of earth embankments under earthquake ground motions calculated from developed predictive GEP-based model (Eq. 8) and also the available relationships (Table 5) are illustrated in Fig. 10. As demonstrated in Figs. 9 and 10, the developed GEP-based model has a high precision compared to the previous relationships for estimation of slope displacement of earth embankments under earthquakes.

It should be noted that convolution of earth embankments behavior under earthquake ground motions has causes to exact not reflecting all influential factors affecting slope displacements in conventional recommendations. However, the available recommendations and relationships are extensively utilized in practical earthquake geotechnical problems. Applying advanced computational approaches such as gene expression programming (GEP) can definitely be an efficacious step to reduce uncertainty in dynamic analysis of earth embankments subjected to earthquake excitations.

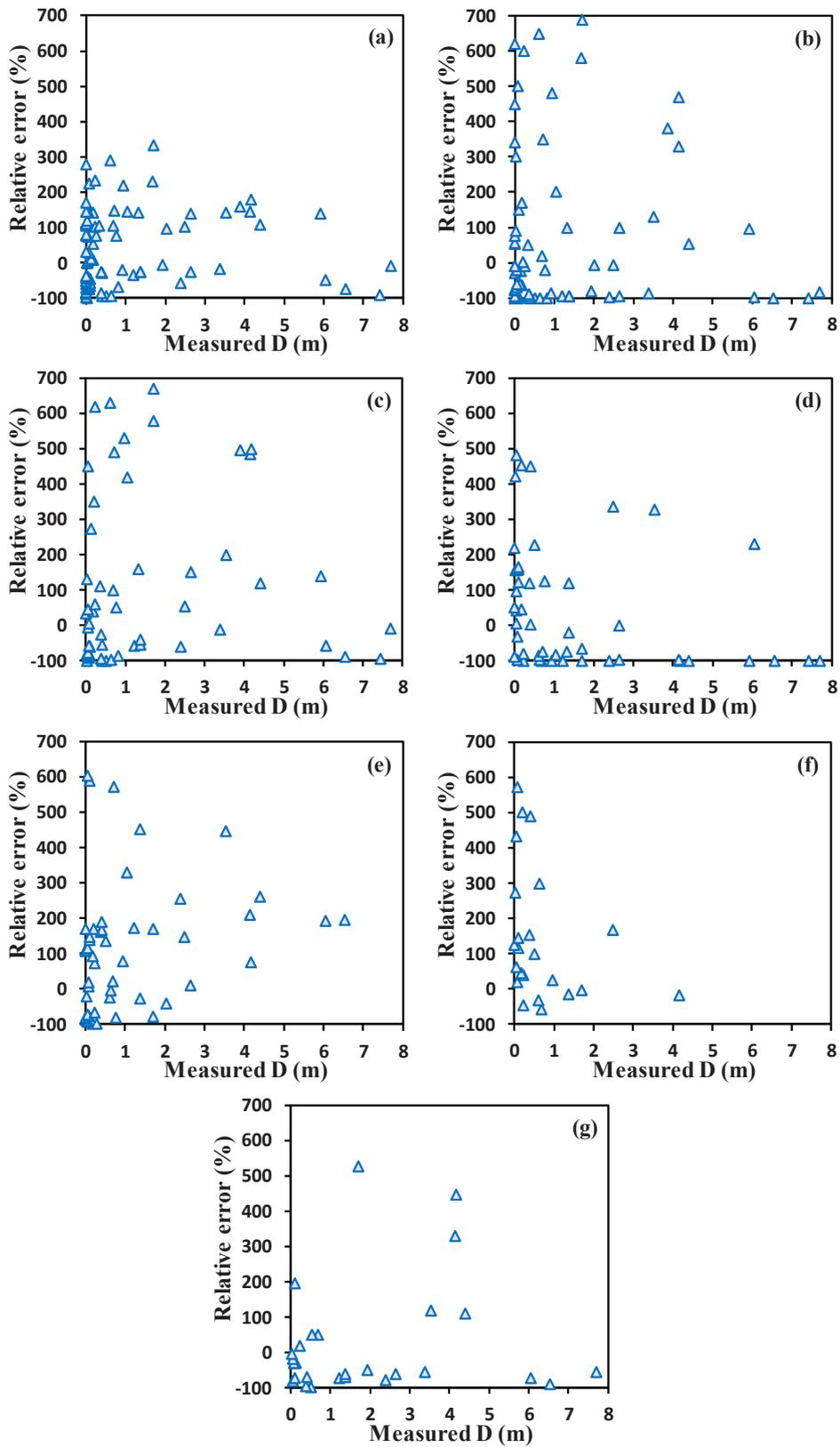

**Fig. 10.** Relative error versus measured seismic displacement, **a)** This study (GEP-based predictive model), **b)** Tsai and Chien [70], **c)** Madiai [41], **d)** Saygili and Rathje [60], **e)** Jibson [34], **f)** Ambraseys and Menu [1], **g)** Hynes-Griffin and Franklin [23].

**Summary and conclusions**

Precise evaluating earthquake induced behavior of earth embankments has a vital role in preliminary seismic analysis of such geostructures. The present study assessed the slope stability of earth embankments under earthquake ground motions. Extensive real case histories of earth embankments under past earthquakes in various parts of the world were collected and analyzed. The most important influential factors that affects the seismic behavior of earth embankments were determined using analysis of the gathered data and available recommendations. The parameters ratio of yield acceleration, $a_y/a_{max}$, earthquake magnitude, $M_w$, and ratio of fundamental period, $T_d/T_p$, are considered as important influential parameters. Using gene expression programming (GEP) a model was developed to predict slope displacement of earth embankments under earthquake ground motions ($D$). Comparison of earthquake induced slope displacement predicted by proposed GEP-based model with the real case histories data demonstrates reasonable accuracy of the model in training stage ($R^2$=0.742, MAE=1.198, RMSE=1.546, SI=0.787, Bias=0.152), validation stage ($R^2$=0.720, MAE=1.675, RMSE=2.115, SI=0.930, Bias=0.476) and all data ($R^2$=0.730, MAE=1.321, RMSE=1.712, SI=0.837, Bias=0.236).

The sensitivity analysis was conducted to study the influence of each parameter on the slope displacements of earth embankments under earthquakes and also to recognize the performance of the developed GEP-based model. The trends of variations of GEP-based predicted ln $D$-$M_w$ under different values of $a_y/a_{max}$ and $T_d/T_p$ were evaluated in comparison with the real case histories of earth embankments. The slope displacement of earth embankments under earthquake ground motion increased by increasing $M_w$ and decreased by increasing $a_y/a_{max}$ and $T_d/T_p$. Investigating variation trend and comparison with real data illustrate the appropriate performance of the proposed GEP-based model in calculation of slope displacement of earth embankments subjected to earthquakes. Finally, the GEP-based predictive model performance was compared with the available relationships for estimation of seismically soil slope displacement. The results clearly confirm higher accuracy of proposed predictive GEP-based model. Certainly, further real cases under different seismic loading could improve the accuracy and performance of predictive models to assess soil slope displacements under earthquake ground motions.

**CRediT authorship contribution statement**

**Siyu Fan:** Writing – original draft, Validation, Project administration, Methodology. **Sanglin Zhao:** Writing – original draft, Methodology, Formal analysis, Datacuration. **Zhenyang Jin:** Writing – original draft, Methodology, Formal analysis, Data curation. **Hamed Javdanian:** Writing – review & editing, Supervision, Methodology.

**Declaration of Competing Interest**

The authors declare that they have no known competing financial interests or personal relationships that could have appeared to influence the work reported in this paper.